\documentclass[preprint]{elsarticle}

\usepackage{lineno,hyperref}

\usepackage{amssymb,amsmath,bm}
\usepackage{booktabs}
\usepackage{multirow}

\journal{arXiv}

\biboptions{authoryear}

\begin{document}

\begin{frontmatter}




\title{The Poisson algorithm: a simple method to simulate stochastic epidemic models with generally distributed residence times}

%
%
\author{Carlos Hernandez-Suarez}

\address{Instituto de Ciencias Tecnolog\'ia e Innovaci\'on\\
Universidad Francisco Gavidia\\
El Progreso St., No. 2748, Colonia Flor Blanca\\
San Salvador, El Salvador\\
cmh1@cornell.edu}

\author{Osval Montesinos-Lopez}

\address{Facultad de Telematica, Universidad de Colima\\
Av. Universidad 333\\
Colima, Colima, 28040, MEXICO\\
oamontes2@hotmail.com}

\author{Ramon Solano-Barajas}

\address{Facultad de Ingenier\'ia Civil, Universidad de Colima\\
Carr. Colima-Coquimatlan, Km 9\\
Coquimatlan, Colima, 28400, MEXICO\\
rsolanob@gmail.com}

\begin{abstract}
Epidemic models are used to analyze the progression or outcome of an epidemic under different control policies like vaccinations, quarantines, lockdowns, use of face-masks, pharmaceutical interventions, etc. When these models accurately represent real-life situations, they may become an important tool in the decision-making process. Among these models, compartmental models are very popular and assume individuals move along a series of compartments that describe their current health status. Nevertheless, these models are mostly Markovian, that is, the time in each compartment follows an exponential distribution. In epidemic models, exponential sojourn times are most of the times unrealistic, for instance, they imply that the probability that a patient will recover from some disease in the next time unit is independent of the time the patient has been sick. This is an important restriction that prevents these models from being widely accepted and trusted by decision-makers. In spite of the need to incorporate algorithms to tackle the problem, literature on the topic is scarce. Here, we introduce a novel approach to simulate general stochastic epidemic models that accepts any distribution for the sojourn times that is efficient. 

\end{abstract}


\end{frontmatter}


\section{Introduction}

A mathematical model is a real-life sketch that allows for experimentation and testing. Engineers have  used models for long time, where temperature, friction, durability and forces play a role in decision making. Similarly, chemists have used mathematical models to analyze chemical reactions with the purpose of optimization, but, despite the long-standing tradition of using mathematical models, their use in biology or medicine is much more recent. Sometimes scientists in these disciplines deal with not well understood phenomena and thus, the set of assumptions is usually large and require the conjunction of different disciplines. From the classical Ross Malaria model \citep{ross1915some} that allowed to conclude that keeping the mosquito population below a threshold would eliminate Malaria, to Blower's findings \citep{Blower1451} that a vaccine with a low level of protective efficacy may do more harm than good by providing a sense of false security to the vaccinated, models may reveal a hidden `cobra effect', that is, something that has not accounted for. 

Mathematical models have been a valuable tool to analyze the demographics of biological populations and have been used to follow the health of populations in the field \citep{caswell2001matrix}. As Cohen  \citep{cohen2004mathematics} stated: ``\textit{Mathematics is biology's next microscope, only better; biology is mathematics' next physics, only better}''.

Epidemics are not only driven by etiological agents but also by the behavior of the population. For this later, mathematical models in epidemiology resemble more a field of economics than of medicine. Examples of  such behaviour may involve the population's reaction to vaccination, abortion, use of face-masks, medication, blood or plasma transfusions, condom use, etc. The fact that a model considers such behaviors does not guarantee that these have been correctly included in the model, which most of the times requires a deep knowledge of the phenomena and interdisciplinary work. For instance, Needle/Srynge Policies (NSOs) that attempt to reduce HIV transmission by providing needles for free, must face the fact that sharing needles is sometimes part of the social experience \citep{kaplan1994circulation}, and also that peer pressure is a determinant factor in acquiring smoking habits \citep{evans1978deterring}.

\begin{figure}[pb]
\begin{center}
\centerline{\includegraphics[width=3in]{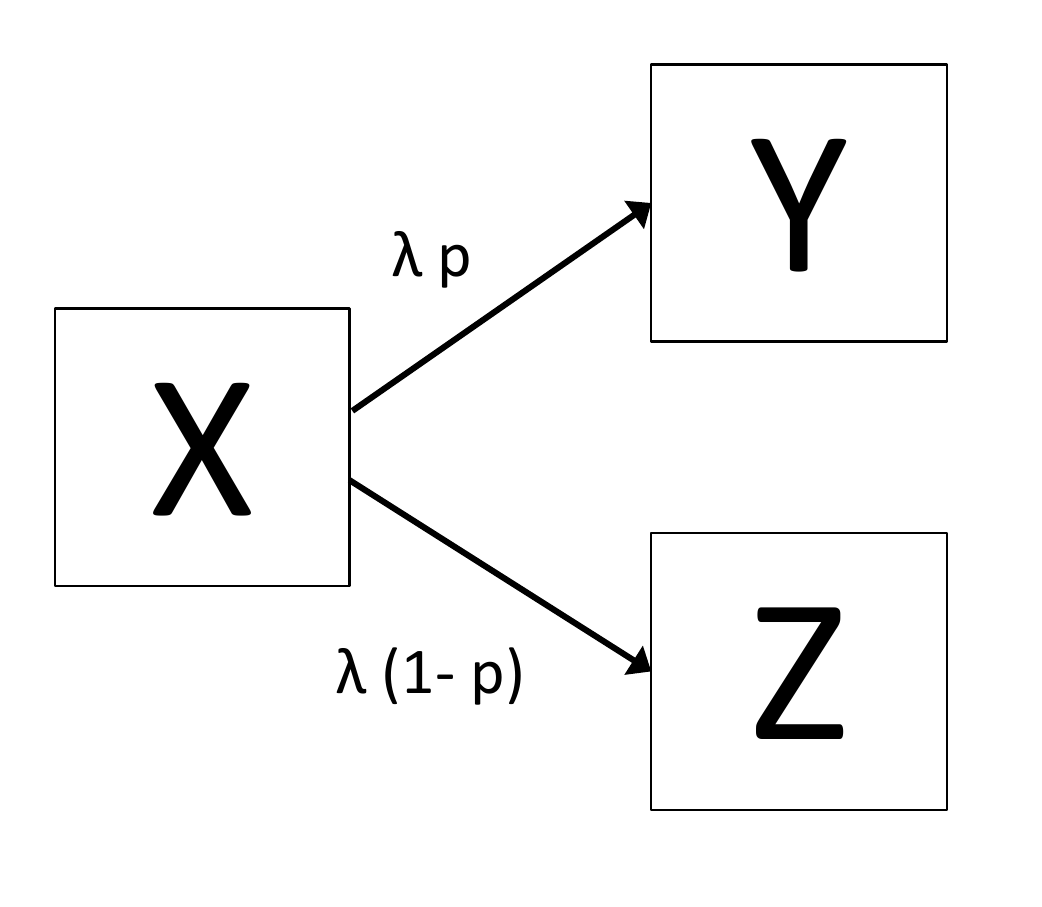}}
\vspace*{8pt}
\caption{The main differences between deterministic and stochastic models can be shown in this Figure. Suppose that at time $t=0$ there is only one individual in compartment $X$. In a deterministic model this individual is \textit{drained} continuously at a rate $\lambda$ and a fraction $p$ of this goes to compartment $Y$, thus, after some finite time $t$, this individual is spread among the three compartments, whereas in a stochastic model the individual is in only one of them.}
\label{Fig1}
\end{center}
\end{figure}

\begin{figure}[pb]
\begin{center}
\centerline{\includegraphics[width=3in]{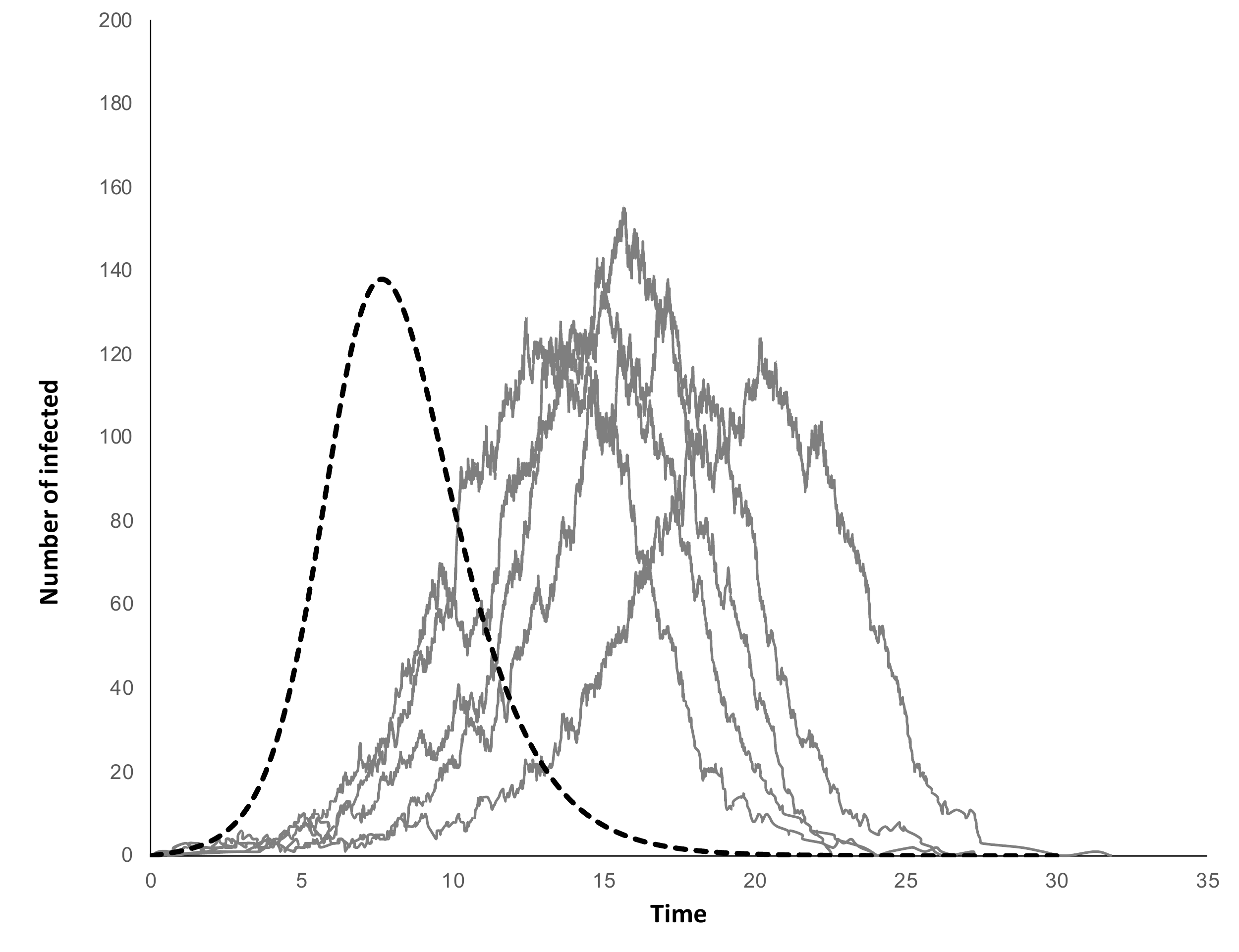}}
\vspace*{8pt}
\caption{Stochastic simulations of as SEIR model. Some simulations vanished quickly and are unnoticeable. The plot shows the number of infectious (I) over time, with $N=1000$ and $R_0=2$. The dashed line is the solution of the deterministic model.}
\label{Fig2}
\end{center}
\end{figure}

The most popular epidemic models consist of a series of stages or compartments that represent the different health status or conditions. Individuals move along the compartments according some specified transition rules. The analysis of these models can be deterministic or stochastic. There are many differences between these two types of models, but here we illustrate the two that are the most relevant. In Figure \ref{Fig1} we can see a stage model with three compartments, $X,Y$ and $Z$. The deterministic model consists of a continuous \textit{drainage} of individuals from all compartments and thus, if at time $t=0$ there is one individual in compartment $X$, after some finite time $t$, the individual will be spread among the three compartments. In contrast, in a stochastic model, the individual can be in only one of them. The most important difference is the source of uncertainty: in deterministic models, based on differential equations, there is no uncertainty once parameters are fixed, whereas in stochastic models, the outcome may vary under the same parameter set. Figure \ref{Fig2} shows 5 stochastic simulations of an SEIR epidemic model shown together with the solution of the deterministic model. Clearly, the distribution of the stochastic simulations is shifted to the right, since small epidemics vanish quickly and are unnoticeable. This will result in differences in the average time to extinction and the epidemics cost (the integral under the curve of infected) and other functionals of time.

\section{Simulations}

Before the advent of computers one could only rely on analytic results which limited the complexity of the models. Computer simulations allow researchers to obtain the response to \textit{`What would happen if...'} in complex situations. In deterministic models, the term \textit{simulation} refers to the numerical solution of a set of differential equations. In stochastic models, it implies the use of pseudo-random numbers to decide when an individual leaves a compartment as well as the origin and destination compartments. Stochastic models allow researchers to obtain information on the variability of the possible outcomes: while in a deterministic model the number of infected at time $t$ is constant, in stochastic models, this number follows a statistical distribution whose moments may be infered by performing a large number of simulations.

In Markovian epidemic models, sojourn times in each compartment are exponential random variables. Simulating these models is easy because the \textit{memoryless property} allows to describe the whole system at any time $t$ with a vector containing the number of individuals in each compartment. But the mean and standard deviation of an exponential distribution are equal, which is a very stringent condition for real-life situations. If sojourn times are not exponential we would need to keep track no only of the number of individuals in any given compartment, but also on the actual time each individual has been there, which affects the decision on who moves next and where to. Keeping this record is computationally intensive.

We first provide a review of the \textit{Gillespie} algorithm \cite{gillespie1976general}, also known as  \textit{time to step} simulation, used in simulating epidemic models with exponential sojourn times and how it is commonly dealt when the sojourn time is not exponential.

\subsection{The \textit{Gillespie} algorithm.}

As mentioned before, whereas in a deterministic model there is a continuous `leaking' between boxes, in the Markovian version of the model, only one transition between two compartments takes place at any time, which is a random event and the time when this transition takes place is another (independent) continuous random variable. The following results from exponential distributions are used by the \textit{Gillespie} algorithm:

\vspace{5mm}

If $X$ and $Y$ are two exponential distributions with respective parameters $\alpha$ and $\beta$, then:

\begin{enumerate}[(i)]

\item
$P(X<Y) = \alpha/(\alpha + \beta)$

\item
The $\min (X,Y)$ follows an exponential distribution with parameters $\alpha+\beta$

\item 
The probability that $X<Y$ is independent of the distribution of $\min (X,Y)$.

\item
If an event occurs at a time which follows an exponential distribution with parameter $\alpha$ but when the event occurs, there is a probability $p$ that event will be of type $A$, then the time to the occurrence of an event of type $A$ follows an exponential distribution with parameter $\alpha p$.
 
\end{enumerate}
 
 Property (iii) implies that the time to the next event is independent of the event that will take place. This is an important property because we can  independently simulate the time to the next event and the event that will take place. Property (iv) is called \emph{thinning} the Poisson processes.
 
Assume that there are $K$ compartments and that transitions at time $t$ from compartment $i$ to compartment $j$ occur a a rate $\delta_{ij}(t)$. For simplification, we will write this rate as $\delta_{ij}$. A stochastic simulation of these models involves: 
 
 \begin{enumerate}

\item
 Calculating the rates of all transitions $\delta_{ij}$.
 \item
 Calculating the total transition rate $R = \sum_{ij} \delta_{ij}$.
 \item
 Calculating the transition probabilities $p_{ij}$ for every possible transition: as $p_{ij} =\delta_{ij}/R$.
 \item
 Simulating when the next transition will take place using an exponential distribution with parameter $R$.
 \item
 Selecting the transition that will take place by sampling the $p_{ij}$'s in step $3$.
 \item
 Updating the system, that is, moving one individual from compartment $i$ to compartment $j$ according the chosen transition in step (5) and adding the simulated time obtained in step (4).
 \item
 Re-starting from step (1) and continuing until $R=0$ or when a stopping time $t_{\max}$ is reached.
 
  \end{enumerate}
 
 Due to property (iii) steps (4) and (5) can be exchanged. For step (4), since the time to the next transition is exponentially distributed, we can simulate it by using $T=-log(r)/R$, where $r$ is a uniform random number $(0,1)$. For step (5), we can select a single sample from a multinomial distribution with parameters $p_{ij}$, for $i=1,2,\ldots,K;j=1,2,\ldots,K$ where the outcome is the index of the event that will take place. 
 
This kind of simulation is also known as \emph{time to event} simulation because we modify the contents in each compartment until the next event takes place. Thus, in a SEIR model the number of times the loop (a)-(g) above has to be executed is about 3 times the number of infected, since each one of them must move eventually through the stages $E$, $I$ and $R$. For $R_0$ large, the number of infected may be close to $N$, the population size.  
 
\subsection{Review of fitting non-exponential sojourn times to the compartments}
 
 As mentioned before, stochastic Markovian models assume that the duration of time in each compartment is exponential because these distributions have the \textit{memoryless property} in which regardless of the time an individual has been in a compartment, the probability that the individual will leave it in the next $s$ units of time is $1-e^{-\lambda s}$, for some $\lambda$. Consequently, it is not necessary to track the current time of an individual in a compartment to decide if it is time to leave it or not. We only need to know how many individuals there are in each compartment and the respective exit rates to decide what happens next. Exponential times implies a constant hazard rate, which is not precisely a characteristic of most stages in most diseases. For instance, an individual that just had surgery may present complications from surgery, and the possibility of a complication in the next unit of time reduces with time, thus, this is an example of a decreasing hazard rate. On the other hand, an individual that has been infected may eventually die or recover eventually and the possibility that one of these events will occur in the next unit of time increases with time, thus, it has a positive hazard rate. 
 
An attempt to approximate a distribution with some mean $\mu$ and some variance $\sigma^2$ can be done using Erlang distributions \citep{anderson1980spread, lloyd2001realistic, wearing2005appropriate, champredon2018equivalence}. This approach is equivalent to divide a compartment into several sub-compartments, each with an exponential duration, so that the sojourn time through all of them has the desired mean and variance. An exact match is sometimes impossible and we need to get as close as we can to the desired moments. There are two problems inherent to using this technique: the first one is that, in a model with $K$ stages (excluding the susceptible state), there are roughly $K-1$ transitions for each individual, and if each of those $K$ stages is divided in $n$ substages, the number of transitions increases by $n$, resulting in more computing time. The second problem is that the approximation of the first two moments of a single stage alone may require a very large number of substages. For instance, if we want to match a distribution with mean 20 days  and standard deviation $\sqrt{8}$, we would need an Erlang distribution with parameters $50$ and $5/2$, that is, $50$ compartments in a row, each with an exponential sojourn time with parameter $5/2$. At some point, the number of compartments may become prohibitive, and some relaxations would be made, and the distribution is only approximated. If this is repeated for a model with several compartments, things can become very complicated.

\section{The Poisson algorithm.}

In an epidemic model, two factors drive transitions between stages or compartments: the pressure for infection and the natural rate at which individuals leave their current stage. Suppose that there is no infection pressure, and that $\bm{t}_{rem}$ is a vector containing the remaining times in the their current stage. Then, the next event will take place exactly in $t_{min}$ units of time, where $t_{min}$ is the minimum value of $\bm{t}_{rem}$. Nevertheless, if there is infection pressure of size $\alpha$, an infection may occur before a transition with probability:

$$
1-e^{-\alpha \ t_{min}}
$$
since the number of events in $(0,t_{min})$ is Poisson with parameter $\alpha \ t_{min}$. Therefore, we can use this to simulate epidemic models with non-exponential sojourn times. Clearly, when sojourn times are exponential, $t_{min}$ is always exponential.

\subsection{A suggested pseudocode to obtain the times of infection \textbf{y}.}
\label{pseudo}

Here we provide a pseudocode to obtain the infection times. Assume there are $K$ compartments in our model (excluding the susceptible compartment) and that there are currently $Z$ individuals that have left the susceptible compartment (the number of accumulated infections so far). Let $\bm{n}$ be a $K \times 1$ vector containing the current number of individuals in each compartment and $\bm{\lambda}$ the corresponding individual infectious rate for each compartment. Let $\bm{T}$ be the $N \times K$ matrix described previously and $\bm{u}$ be a $Z \times 1$ vector with the remaining excess life of those individuals that have left the susceptible compartment. Let $\bm{s}$ be a vector with the corresponding actual stage for each individual with the corresponding excess life in $u$.

Suppose that the model has $K$ stages. \textit{event} occurs when an individual moves from one compartment to another. Start initializing a vector $\bm{y}$ and a matrix $\bm{W}$. Vector $\bm{y}$ will store the times when \textit{events} occur whereas $\bm{W}$ stores the respective number of individuals in each compartment. Usually, at the beginning: $\bm{y}=[0]$ and the in initial size of $\bm{W}$ is $1 \times K$:
$$
\bm{W}=[N-1,1,0,0,0 \ldots 0]
$$

Also let:

\begin{itemize}
\item
$\lambda$, a vector of size $K \times 1$ containing the individual infection rates for individuals in every stage.

\item
$\bm{s}$, the current stage of every individual that has left the susceptible compartment.
\item
$\bm{t}_{rem}$, a vector containing the remaining time in its current stage for all this in individuals tha have left the susceptible compartment.
\item
$\bm{n}$, a vector of size $K \times 1$ with the current number of individuals in each compartment.  Observe $\bm{n}$ is the last row of $\bm{W}$.

\end{itemize}

The pseudo-algorithm goes as follows:\\

\begin{enumerate}[i)]
\item
Find $u^*$, the minimum value in $\bm{t}_{rem}$. Let $r$ be the index of this individual.
\item

Calculate the total infection pressure on the susceptible, $\alpha = \bm{n}' \bm{\lambda}(S/N)$, where $S$ is the current number of susceptible.

\item
Simulate $Y$, an exponential distribution with parameter $\alpha$. Set $\bm{t}_{rem}=\bm{t}_{rem}-Y$

\item
if $Y < u^*$, an infection occurs, in this case:
\begin{itemize}

\item
Add a new row to $\bm{s}$, set this equal to the stage where susceptible move when infected.
\item
Add a new row to $\bm{t}_{rem}$, set this equal to $x$, a simulation of the distribution of the stage when susceptible move when infected.
\item
Update $n$. Append $Y$ to vector $\bm{y}$.
\end{itemize}

if $Y > u^*$, the $r$-th individual moves to the next stage:
\begin{itemize}
\item
$\bm{s}(r)=\bm{s}(r)+1$
\item

Let $\bm{t}_{rem}(r) =x$, a simulation of the distribution of the current stage of this individual. If the individual has arrived to a stage where it will stay forever, set $\bm{t}_{rem}(r)=\infty$. 

\end{itemize}

\item
Update $\bm{n}$ and append this to $\bm{W}$. Append $u^*$ to vector $\bm{y}$.

\item
Repeat from (i).

\end{enumerate}

\section{Examples}

We can classify epidemic compartmental models in two categories: models in which no stage can be visited more than once, and models in which one or more stages can be visited more than once. The difference is that in the first case, there is a maximum for the number of transitions that an individual can make, thus, the size of matrix $\bm{T}$ containing the duration in each stage can be preset to $N \times K$. In the second case the matrix can be of any size. Here, we will deal only with the fist case and indicate how to proceed in the second case.

\subsection{Example 1: the SEIQR model}

\begin{figure}[ht]
\begin{center}
\includegraphics[width=4.7in]{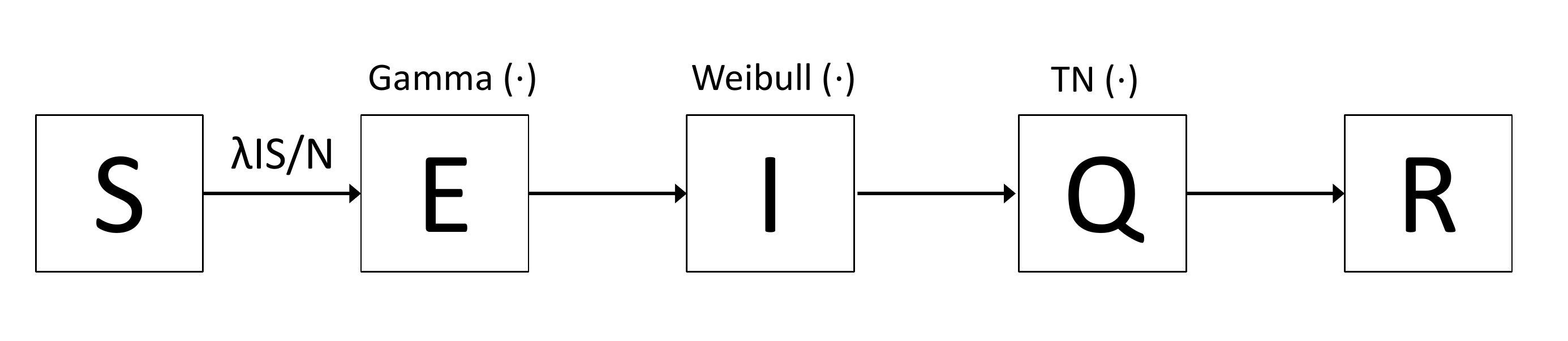}
\caption{A Susceptible- Latent- Infectious- Quarantined- Removed (\emph{SEIQR}) model. The notation on top of each stage indicates the associated statistical distribution.}
\label{Fig6}
\end{center}
\end{figure}

The model in Figure \ref{Fig6} is a Susceptible- Latent- Infectious- Quarantined- Removed (\emph{SEIQR}) model. Infectious individuals have contacts according a Poisson process with parameter $\lambda$ and encounter susceptibles with probability $S/N$, thus, individuals leave the $S$ compartment at a rate $\lambda I S /N$. Once they leave the $S$ compartment they go through the rest of stages $E,I,Q$ and finally arrive to $R$, where they remain. The notation on top of each stage indicates the associated statistical distribution. For this model, the infectious contact rates are: $\bm{\lambda}'=[0,2.1,0]$ for stages $E,I$ and $Q$ respectively.

\begin{figure}[ht]
\begin{center}
\includegraphics[width=4.7in]{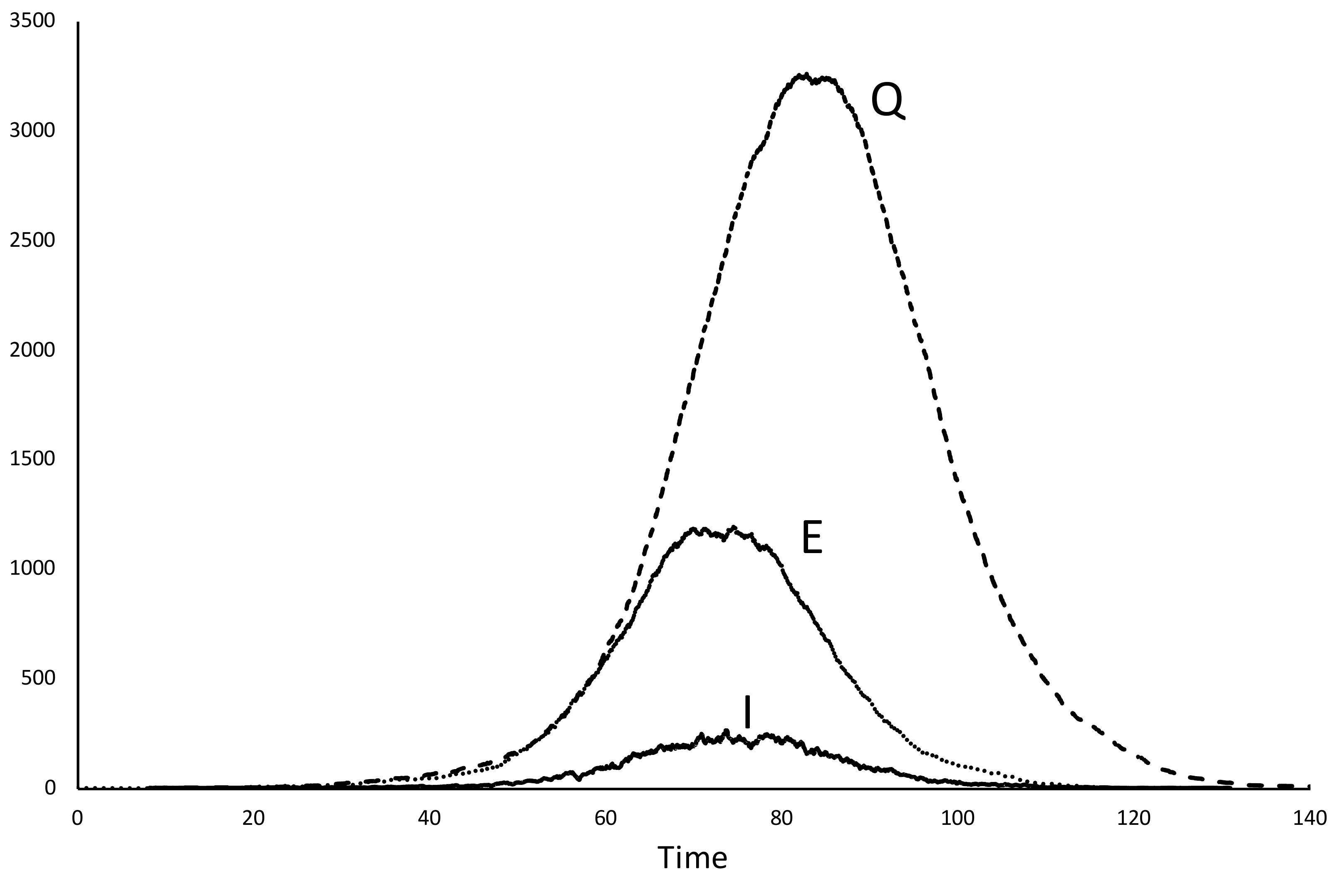}
\caption{A simulation of the \textit{SEIQR} model of Figure \ref{Fig6}.}
\label{Fig7}
\end{center}
\end{figure}

In this case we use a Gamma$(2.3, 2)$ distribution for the $E$ stage, a Weibull $(1,2)$ for the $I$ stage and a Gamma $(5.7, 2.5)$ for the $Q$ stage. Figure \ref{Fig7} shows a single simulation of the \emph{SEIQR} for $N=10,000$. 

\subsection{Example 2: a more general model}

\begin{figure}[ht]
\begin{center}
\includegraphics[width=4.7in]{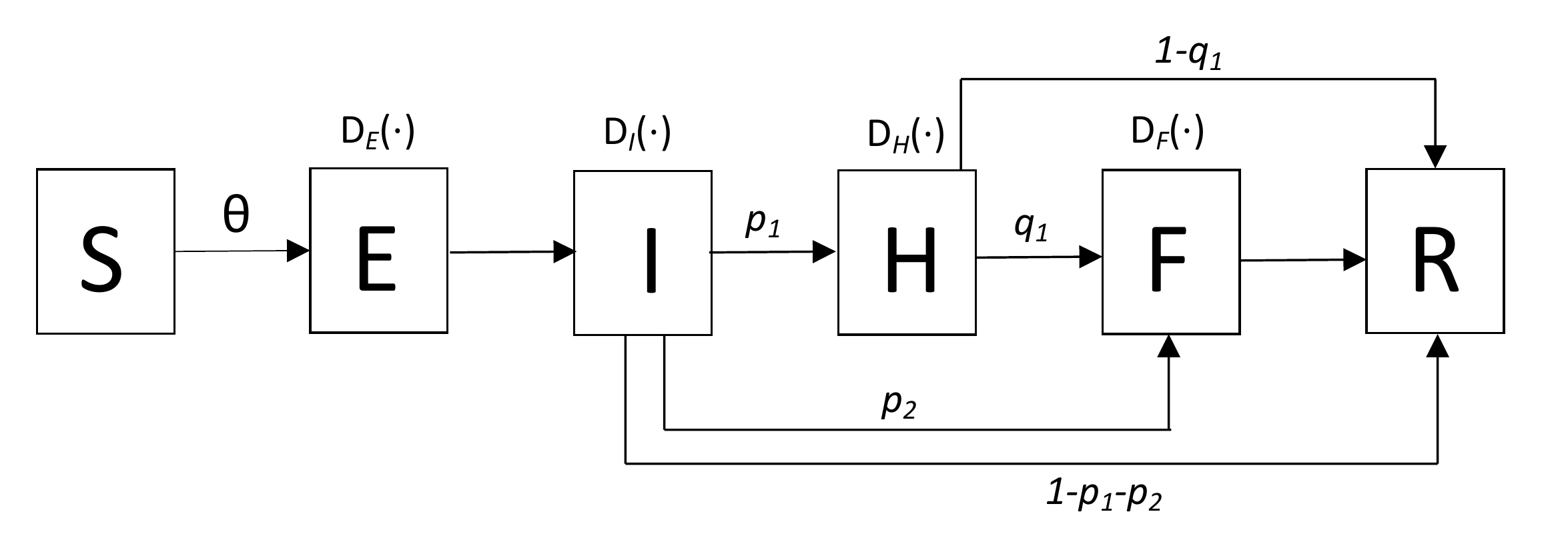}
\caption{An Ebola transmission model adapted from Legrand et al. (2007), in which hospitalized and unburied are a source of infection. \textit{S}, Susceptible individuals; \textit{E}, Exposed individuals; \textit{I}, Infectious; \textit{H}, hospitalized; \textit{F}, dead but not yet buried; \textit{R}, removed. $Di(\cdot)$ indicates some general distribution for the duration in stage $i$ described in Table \ref{tab:2}. The total infection rate is $\Theta = (\lambda_I I + \lambda_H H + \lambda_F F)S/N$.}
\label{Fig8}
\end{center}
\end{figure}

The  model depicted in Figure \ref{Fig8} is an example where individuals are infectious at different stages and with different degree of infectiousness. This is an Ebola model where infectious individuals are infectious even while hospitalized or even dead before being buried, \citep{legrand2007understanding}. This is a complete example for the models in this category since not all stages may be visited and there are several kinds of infectious stages with differential degree of infectiousness. Here, the model has been modified to admit any distribution for the duration in stage $i$, $D_i(\cdot)$. 

We start by generating a matrix \textbf{T} with four columns, one for each of the stages $E,I,H$ and $F$. Then, there are several strategies to simulate the possible paths that an individual can follow and here we chose a simple approach. Table \ref{paths} shows a list with the possible visited stages and the associated probabilities. We can simulate from this distribution and generate $N$ possible paths which can then be recorded in a matrix \textbf{P} of $0$'s and $1$'s depending on the stages $\{E,I,H,F\}$ being visited or not.

\begin{table}[ht]
\centering
\caption{Possible paths after infection with their probabilities.}
\begin{tabular}{@{}cc@{}}\\
\toprule
Path    & Probability   \\ \midrule
$E-I-H-F$ & $p_1 q_1$     \\
$E-I-H$   & $p_1 (1-q_1)$ \\
$E-I-F$   & $p_2$         \\
$E-I $    & $1-p_1-p_2$   \\ \bottomrule
\end{tabular}
\label{paths}
\end{table}
After the simulation of the $N$ paths, the resulting matrix will look like:

 \begin{equation}
\mathbf{P}=
\begin{pmatrix}
1 & 1 & 1 &0 \\
1 & 1 & 0 &1 \\
1 & 1 & 0 &0 \\
 & & \vdots & &\\
1 & 1 & 1 &1 \\
1 & 1 & 0 &0 
\end{pmatrix}\ \ \ \
\label{P}
\end{equation}
\noindent
and then matrix \textbf{T} can be redefined as:

\begin{equation}
 \bf{T}= \bf{T} \oslash \bf{P}
 \label{T}
 \end{equation}
matrix $\bm{T}$ contains the time spent for each individual in each state, and we can apply the pseudocode suggested in \S \ref{pseudo} to this matrix. The first five rows of $\bm{T}$ are:

$$
\bf{T}=\bordermatrix{
& \textit{E} & \textit{I} & \textit{H} & \textit{F} \cr
& 5.145 & 4.239  & 14.807 & 0 \cr
& 4.048 & 2.729  & 0  & 2.000 \cr
& 8.279 & 10.502 & 32.745 & 2.000 \cr
& 5.283 & 3.645  & 21.114 & 2.000 \cr
& 6.143 & 5.695  & 10.694 & 0\cr}
$$
where we can see, for instance, that the first and fifth individuals were buried immediately, and the second died without being hospitalized.

\subsubsection{Example} 

We performed a single simulation of the Ebola model \citep{legrand2007understanding} to illustrate this using arbitrary parameters and distributions for the duration on every state. For the model in Figure \ref{Fig8}, we use $\lambda_I=0.15, \lambda_H = 0.1, \lambda_F = 0.05, p_1=0.797, p_2=0.163,q_1=0.9$, whereas we use the distributions indicated in Table \ref{tab:2} for each state. The infectious contact rates are: $\bm{\lambda}'=[0. ,0.15, 0.1,  0.05]$ for stages $E,I,H$ and $F$ respectively and use $N=10,000$.  The Python code for this example is provided also as Supplementary material.

\begin{table}[ht]
\centering
\caption{The distributions used to simulate the Ebola model from \cite{legrand2007understanding}  }
\begin{tabular}{@{}ccccc@{}}\\
\toprule
Stage & Distribution                   & Parameters        & Mean & SD   \\ \midrule
$E$     & Truncated Normal in {[}4,10{]} & $\mu=7,\sigma=3$, & 7    & 1.62 \\
$I$     & Rayleigh                       & $4.3$             & 5.39 & 2.81 \\
$H$     & Gamma                       & $\alpha=7,\beta=2.1$         & 14.7 & 5.55  \\
$F$     & Constant                       & 2                 & 2    & 0    \\ \bottomrule
\end{tabular}
\label{tab:2}
\end{table}

\begin{figure}[ht]
\begin{center}
\includegraphics[width=4.7in]{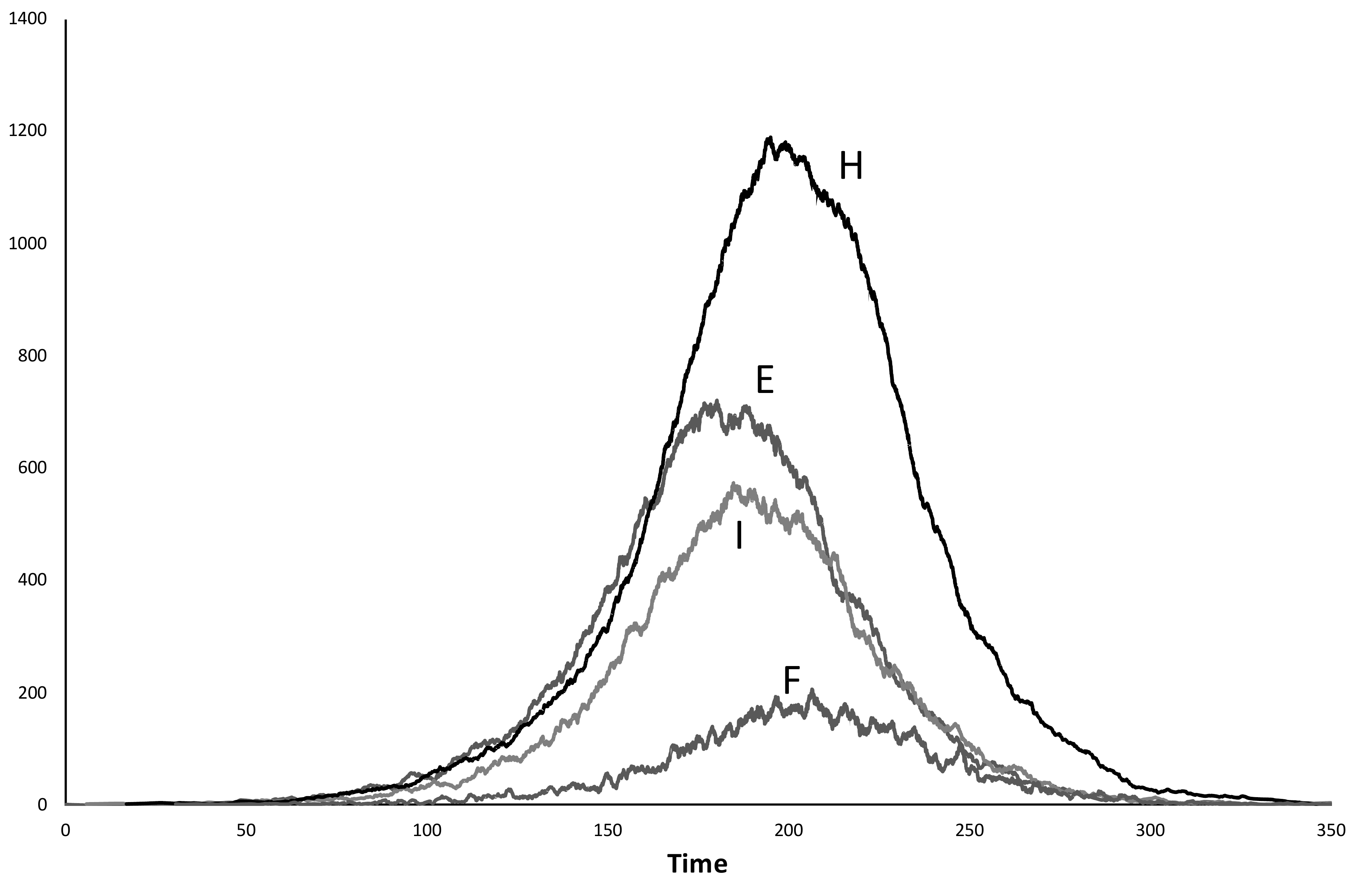}
\caption{A single simulation of an Ebola epidemics depicted in Figure \ref{Fig8} in which hospitalized and unburied are a source of infection. The duration in each stage is described in Table \ref{tab:2}.}
\label{Fig9}
\end{center}
\end{figure}

\section{Discussion}

The purpose of this research is to introduce a method to simulate stochastic epidemics when sojourn times follow a general distribution. Using the true distribution may be useful to obtain better estimates of functionals of path integrals, for instance, to estimate the total bed-days required in hospitalization or in IC units.

Some methods have been suggested to deal with the problem of non-exponential sojourn times, for instance, the \emph{event-driven algorithm} \cite{kiss2017mathematics}. This algorithm suggests building a list of all possible events and make a \emph{priority queue} where elements are continuously added and sorted. Since sorting is computationally intensive, the procedure may be very inefficient. For instance, in an \emph{SIR} model, with $I$ infectious, we need to build a list with the indexes of the future contacts for each of the $I$ infectious. We also need a list with the timings for those contacts. The next infection is the contact that will occur sooner, provided the individual is a susceptible. New infections compete in the infection process thus for every new infection we need to generate a list for future contacts of this newly infected, as well as their timings. In addition, each new infection requires removing its index from the list of future contacts of other infected. It is difficult to compare how efficient is the Poisson algorithm against the \emph{priority queue} algorithm whose purpose is to get the job done, that is, simulation of non-exponential sojourn times, and not to be efficient. The \emph{event-driven algorithm} may become incredible complicated in complex epidemic models, involving more than three categories.

There is no benchmark to compare the Poisson algorithm, so we chose as a reference the time that would take to simulate the same model under exponential sojourn times, using the  \textit{Gillespie} algorithm. Table \ref{comparison} shows a comparison of the \textit{Gillespie} algorithm and ours, for three epidemic models: the basic \emph{SIR} model, the \emph{SEIQR} model and the \emph{SEIHFR} Ebola model we used before. We use the Poisson algorithm with both an exponential distribution and a general distribution. To reduce the frequency of simulations ending with a small epidemics, we started with $I_0=10$ initial infected individuals. Computer times correspond to an 2.7 GHz MacBook Pro.

\begin{table}[ht]
\centering
\caption{Algorithm comparison. Averages from $100$ simulations.}
\begin{tabular}{@{}cccc@{}}\\
\toprule
& \multicolumn{3}{c}{Average time per simulation (s) ($N=10^4, 10^5)$ }  \\ \cmidrule {2-4}
Model & Gillespie                   & Poisson ($E^*$)   &   Poisson ($G^\dagger$)    \\ \midrule
\emph{SIR}     & $(0.07,0.66)$ & $(0.42,4.87)$ & $(0.43,5.42)$   \\
\emph{SEIQR}     & $(1.98,19.6)$ & $(1.22,23.3)$ &  $(1.26,24.1)$  \\
\emph{SEIHFR}     & $(2.96,29.58)$ & $(1.31,17.74)$ & $(1.52,20.1)$  \\  \bottomrule
\multicolumn{4}{l}{ $^*$ Exponential sojourn time.} \\
\multicolumn{4}{l}{ $^\dagger$ Generalized sojourn time.} \\

\end{tabular}
\footnotetext{hello}
\label{comparison}
\end{table}

Table \ref{comparison} suggests that the Poisson algorithm outperforms the \textit{Gillespie} algorithm when models are complex, where it is even faster than the \textit{Gillespie} algorithm under exponential sojourn times. It can also be seen that the difference between the Poisson algorithm using an exponential o a general sojourn time, is minimal. Recall this comparison must be considered with caution since the use of the Poisson or \emph{Gillespie} algorithm depends on the distribution we are assuming for the sojourn time.

An important advantage of the Poisson algorithm is that some functionals of the epidemic process can be more accurately simulated, which allows to construct better point and interval estimates. Consider for instance the hospital bed days, defined as the sum of the units of time spent by all individuals that where hospitalized. Recall exponential distributions have the same mean and standard deviation, therefore an hospitalized individual with departure rate from hospital of $14.7^{-1}$ will last on average $14.7$ days hospitalized with an S.D. of $14.7$ days, whereas if the duration follows a Gamma$(7, 2.1)$ distribution the average duration is the same but has a standard deviation of $5.55$ days, that is, Exponential sojourn times are in this particular case $2.6$ times larger, which is also the amount in which confidence intervals increase. Similarly, there may be compartments with constant duration, as most quarantine times or post-surgery observation times, and these models are resistant to the \emph{Gillepsie} algorithm, that require exponential sojourn-times in all stages.

Observe that models in which individuals move between compartments $i$ and $j$ according to some fixed probability $p_{ij}$, implicitly consider that this probability is constant and independent of the sojourn times. It is reasonable to assume that $p_{ij}$ was calculated using a frequentist approach, as it is customary in Markov models wit $p_{ij}= n_{ij}/n_i$ where $n_i$ is the number of individuals that visited stage $i$ and $n_{ij}$ is the number of transitions between from stage $i$ to $j$. If it happens that this probability depends on the length of stay in stage $i$ the transitions are not Markovian anymore. For instance, the longer an individual has been in a recovery room after a surgery, the less likely is that the individual will present post-intervention complications. An important advantage of the Poisson algorithm presented here is that it is possible to include a relaxation in the transition probabilities between compartments. For instance, in the Ebola model in Figure \ref{Fig8}, suppose that the probability that an infected individual moves from stage $H$ (hospitalized) to stage $F$ (dead) increases the longer the individual stays in the $H$ compartment, for example, following the relationship $e^{-\alpha t}$ where $t$ is the time spent in stage $H$, then, we can construct vector $\mathbf{p}$ as a function of the residence times in stage $H$, $\mathbf{t}_H$:

$$
\mathbf{p} = e^{-\alpha \mathbf{t}}
$$

thus, building the matrix $\bm{P}$ in (\ref{P}) with these considerations. Clearly, each row of $\bm{P}$ must be generated with information on the particular values of $p_1,p_2$ and $q_1$ for each individual. The possibility of making individual transition probabilities opens the door to more realistic models.

\end{document}